\documentclass{iaus}
\usepackage{graphicx}

\title[Photoionization models of the Eskimo nebula] 
{Photoionization models of the Eskimo nebula: evidence for a binary
central star?}

\author[A.~Danehkar, D.\,J.~Frew, Q.\,A.~Parker \& O.~De\,Marco]   
{A.~Danehkar$^1$, D.\,J.~Frew$^1$, Q.\,A.~Parker$^{1,2}$
 \and O.~De\,Marco$^1$}

\affiliation{$^1$Department of Physics and Astronomy, Macquarie University, Sydney, NSW 2109, Australia \\[\affilskip]
$^2$Australian Astronomical Observatory, PO Box 296, Epping, NSW
1710, Australia\\email: {\tt ashkbiz.danehkar@mq.edu.au;
david.frew@mq.edu.au}}

\pubyear{2011}
\volume{282}  
\pagerange{119--126}
 \jname{From Interacting Binaries to
Exoplanets: Essential Modeling Tools} \editors{Mercedes T. Richards \&
Ivan Hubeny, eds.}
\begin{document}

\maketitle

\begin{abstract}
The ionizing star of the planetary nebula NGC 2392 is too cool to
explain the high excitation of the nebular shell, and an additional
ionizing source is necessary.  We use photoionization modeling to estimate
the temperature and luminosity of the putative companion.
Our results show it is likely to be a very hot ($T_{\rm eff}$ $\simeq$ 250\,kK),
dense white dwarf. If the stars form a close binary, they may merge within a Hubble
time, possibly producing a Type Ia supernova.
\end{abstract}

\firstsection 
\section{Introduction}

NGC 2392 is a bright, double-envelope planetary nebula (PN), nicknamed the
Eskimo nebula, with a bright hydrogen-rich
central star (CSPN). The effective
temperature, derived from
spectral-line fitting, is 43\,kK (\cite[M\'{e}ndez et al.
2011]{Mendez_etal11}). However, the surrounding PN exhibits
high-excitation emission lines, such as He II $\lambda$4686
and [Ne \textsc{v}] $\lambda$3426, which cannot be produced by the visible
star. In particular, the presence of [Ne \textsc{v}]
implies $T_{\rm eff} > 100$\,kK for the ionizing source. It
seems that a hot companion is needed to supply
the hard-UV photons, likely to be a white dwarf.  If
this is the case, the Eskimo will be a valuable
addition to the small sample of PNe with (pre-)double-degenerate nuclei.

In this work, we aim to estimate the luminosity, temperature and
mass of the optically invisible secondary star in NGC 2392 through
photoionization modeling.  We use the 3-D code
MOCASSIN (\cite[Ercolano et al. 2003]{Ercolano_etal03}) to model the
PN emission lines. We also investigate an alternative hypothesis, i.e.
that the high-excitation lines are due to shocks produced by a fast
bipolar outflow from the CSPN. We use the 1-D shock
ionization code Mappings-III (\cite[Sutherland \& Dopita
1993]{SutherlandDopita93}) to test the feasibility of this
hypothesis.

\section{Modeling}

We initially adopted the observed nebular line intensities and
abundances (except log ${\rm S}/{\rm H}=-5.16$) from \cite[Pottasch
et al. (2008)]{Pottasch_etal08}, and used plasma diagnostics to
derive the electron temperature and density in the usual way. We
adopted a distance of 1.8\,kpc following Pottasch et al. (2011). As
expected, the photoionization modeling showed it was necessary to
use a model with a heterogeneous density distribution.  This was
constructed from narrow-band images and kinematic data following
\cite[O'Dell et al. (1990)]{ODell_etal90}, and we adopted densities
of 3000 and 1300 cm$^{-3}$ for the inner prolate spheroid and outer
zone, respectively.  We used NLTE model atmosphere fluxes from the
grid of \cite[Rauch (2003)]{Rauch_03}. Our first attempt to
determine the characteristics of the putative companion shows that a
very hot, high-gravity white dwarf  with $T_{\rm eff} = 250$\,kK and
$L/L_{\odot} = 650$ is a plausible source of the additional ionizing
photons.   We compare our results with the observed spectrum in
Table 1, and compare the model output to the $[$O \textsc{iii}$]$
image in Figure 1.


\begin{table}[tbp]
\caption{Photoionization model output versus observations.} \label{table_lines}
\begin{center}
{\small
\begin{tabular}{lccc|lccc}
\hline Ion & $\lambda$({\AA}) & Obs. & Mod.~~~ & ~~~Ion &
$\lambda$({\AA}) & Obs. & Mod. \\
\hline $[$Ne \textsc{v}$]$ & 3426 & 4.0 & 2.3 &
~~~$[$N \textsc{ii}$]$ & 5755 & 1.6 & 2.6  \\
$[$O \textsc{ii}$]$ & 3727 & 110&  107  &
~~~He \textsc{i} & 5876 & 7.4 & 7.5  \\
$[$Ne \textsc{iii}$]$ & 3869 & 105 & 130   &
~~~H$\alpha$ & 6563 & 285 &   282\\
H$\gamma$ & 4340 & 47  & 47   &
~~~$[$N \textsc{ii}$]$  & 6584 & 92  & 129 \\
$[$O \textsc{iii}$]$ & 4363 & 19 & 13   &
~~~$[$S \textsc{ii}$]$  & 6717 & 6.7 &  3.2 \\
He \textsc{ii} & 4686 & 37 &   35 &
~~~$[$S \textsc{ii}$]$ & 6731 & 8.6 & 4.6  \\
H$\beta$ & 4861 & 100 & 100 &
~~~$[$Ar \textsc{iii}$]$ & 7135 & 14 &  12 \\
$[$O \textsc{iii}$]$ & 5007 & 1150 & 1143   &
~~~$[$S \textsc{iii}$]$  & 9532 & 91 &  94  \\
\hline
\end{tabular}
}
\end{center}
\end{table}

While various shock models can roughly reproduce the [Ne \textsc{v}]
flux, they fail to reproduce the ionization structure of the other
lines.  Our photoionization model predictions generally agree with
the observations, adopting a volume filling factor of 0.07 (see \cite[Boffi \&
Stanghellini 1994]{BoffiandStanghellini94}).
We note that \cite[Guerrero et al. (2011)]{Guerrero_etal11} have
discovered a very hard X-ray source coincident with the CSPN.   If
the stars form a close binary (see \cite[M\'{e}ndez et al.
2011]{Mendez_etal11}), it is possible that the X-rays are produced
from accretion of material on to the companion. The high effective
temperature could also be explained by re-heating of this star.  We
also estimate the stellar masses from evolutionary tracks. The WD
mass is $\approx$$1\,M_{\odot}$, and the total mass is
$\approx$$1.6\, M_{\odot}$, which exceeds the Chandrasekhar limit.
Hence, the system is a potential Type Ia SN progenitor. Further
observations are needed to better understand the nature of this very
interesting system.

\begin{figure}[tb]
\begin{center}
\includegraphics[width=4.5in ]
{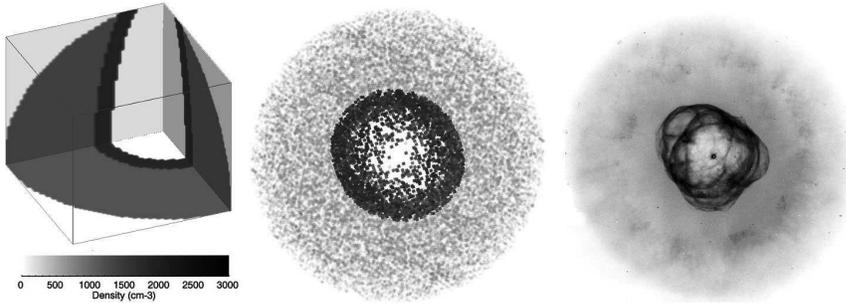}
\end{center}
\caption{(Left) Cross-section of the density distribution
used for NGC~2392. (Right) Computed surface brightness of NGC 2392 in $[$O
\textsc{iii}$]$ $\lambda$5007 compared with the $HST$ image. } \label{fig_pn_hrd}
\end{figure}

\section*{Acknowledgments}
AD acknowledges receipt of an MQRES PhD Scholarship and an IAU Travel Grant.

\end{document}